\documentclass[%
 aip,
 amsmath,amssymb,
 reprint,%
]{revtex4-1}

\usepackage{graphicx}
\usepackage{dcolumn}
\usepackage{bm}
\usepackage{amsmath}
\usepackage{braket}
\usepackage[colorlinks,allcolors=cyan]{hyperref}
\usepackage[utf8]{inputenc}
\usepackage[T1]{fontenc}
\usepackage{mathptmx}
\usepackage{etoolbox}

\makeatletter
\def\@email#1#2{%
 \endgroup
 \patchcmd{\titleblock@produce}
  {\frontmatter@RRAPformat}
  {\frontmatter@RRAPformat{\produce@RRAP{*#1\href{mailto:#2}{#2}}}\frontmatter@RRAPformat}
  {}{}
}%
\makeatother
\begin{document}

\preprint{}

\title[]{Harnessing collective radiative phenomena on a photonic kagome lattice}

\author{Ignacio Salinas}
\author{Javier Cubillos Cornejo}%
\affiliation{ 
Departamento de F\'isica, Facultad de Ciencias F\'isicas y Matem\'aticas, Universidad de Chile, Chile}
\affiliation{Millenium Institute for Research in Optics - MIRO}

\author{Alexander Szameit}
\affiliation{Institute for Physics, University of Rostock, Albert-Einstein-Strasse 23, 18059 Rostock, Germany}

\author{Pablo Solano}
\affiliation{Departamento de F\'isica, Facultad de Ciencias F\'isicas y Matem\'aticas, Universidad de Concepción, Concepci\'on, Chile}
\affiliation{CIFAR Azrieli Global Scholars program, CIFAR, Toronto, Canada}

\author{Rodrigo A. Vicencio}%
\email{rvicencio@uchile.cl}
\affiliation{ 
Departamento de F\'isica, Facultad de Ciencias F\'isicas y Matem\'aticas, Universidad de Chile, Chile}
\affiliation{Millenium Institute for Research in Optics - MIRO}

\date{\today}

\begin{abstract}
Photonic lattices enable experimental exploration of transport and localization phenomena, two of the mayor goals in physics and technology. In particular, the optical excitation of some lattice sites which evanescently couple to a lattice array emulates radiation processes into structured reservoirs, a fundamental subject in quantum optics. Moreover, the simultaneous excitation of two sites simulates collective phenomena, leading to phase-controlled enhanced or suppressed radiation, namely super and subradiance. This work presents an experimental study of collective radiative processes on a photonic kagome lattice. A single or simultaneous -- in or out-of-phase -- excitation of the outlying sites controls the radiation dynamics. Specifically, we demonstrate a controlable transition between a fully localized profile at the two outlying sites and a completely dispersed state into the quasi-continuum. Our result presents photonic lattices as a platform to emulate and experimentally explore quantum optical phenomena in two-dimensional structured reservoirs, while harnessing such phenomena for controlling transport dynamics and implementing all-optical switching devices.
\end{abstract}

\maketitle

\section{\label{intro}Introduction}

Injecting light on an impurity site excites a non-bounded mode, which radiates energy into a given lattice~\cite{impu1,ussub23}. This phenomenon is analog to an atom radiating into a structured reservoir, a fundamental problem in quantum optics \cite{Lambropoulos2000}. For a weakly coupled lattice impurity, the system mimics the radiation of an atom into a continuum~\cite{LuisQuantum,Longhi06,Longhi08,Longhi08prl,Crespi2019,longhi21}, where the decaying dynamics is primarily exponential with a slow power-law decay at longer times ~\cite{Crespi2019}. In contrast, strongly coupled impurities lead to hybrid atom-photon bound states \cite{Lambropoulos2000,Lombardo2014,Sanchez-Burillo2017}. The coupling of two or more impurities to the same lattice reproduces the collective dynamics of many atoms interacting with a common reservoir in the single-photon regime, leading to super and subradiance behavior \cite{John1995,Yang2013,Hood2016,Gonzalez-Tudela2017,Gonzalez-Tudela2017b,Yu2019}. Consequently, photonics lattices offer the potential to study novel quantum optical effects in otherwise typically inaccessible regimes, such as delayed-induced non-Markovianity \cite{Pablo2020,longhi20}, topological reservoirs \cite{Bello2019}, or exploring radiation phenomena in two-dimensional (2D) structured reservoirs. 

The kagome lattice is historically known as the most frustrated 2D system in magnetism due to the impossibility of forming an antiferromagnetic state~\cite{atwood}. Also, this lattice allows studying the interaction between topology and correlations~\cite{guimire} due to the coexistence of Dirac cones and a Flat Band (FB). The first theoretical study of its photonic implementation searched for localized nonlinear cubic solutions outside of the bands ~\cite{Law1}, without a special focus on the linear properties of this lattice. Then, a study on 2D discrete nonlinear dynamics showed the possibility for the mobility of highly compact nonlinear solutions~\cite{kag1}, something that was indeed forbidden for standard nonlinear Kerr 2D systems~\cite{rep1}. A photonic kagome lattice was also suggested for non-diffracting image transmission based on the coherent linear combination of FB states~\cite{kag2}. 
Photonic kagome lattices have been fabricated by diverse means~\cite{Denz1,Wang1,Jo1}, by using photorefractive SBN crystals~\cite{Chenkag1,Chenkag2} or femtosecond (fs) laser written structures~\cite{Alex1}. However, previous experiments were limited to lattices with only few lattice sites~\cite{Hassan,cornerChen}. Moreover, the intrinsic ellipticity of the fs technique produces non-symmetric coupling constants and, for example, the FB properties of a geometrically symmetric kagome lattice~\cite{fbChen,repFB} are simply lost, transforming the system into a Graphene-like structure~\cite{Plotnik1}, already studied in diverse contexts of physics~\cite{gra1}.

\begin{figure}[t!]
\centering
\includegraphics[width=0.99\linewidth]{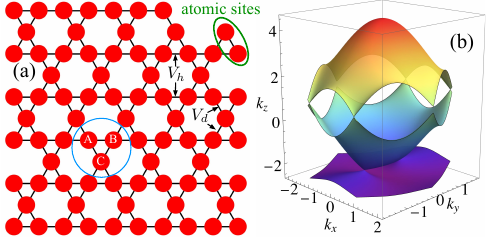}
\caption{(a) A kagome lattice. (b) Linear spectrum for $V_d/V_h=1.2$.}
\label{fig1}
\end{figure}

In this work, we study radiation phenomena on a photonic kagome lattice evanescently coupled to two outlying sites that emulate two radiating atoms. We numerically and experimentally demonstrate that the optical excitation of the outlying sites produces a radiation pattern into the lattice that initially decays exponentially with plateaus at around one-half of their initial energy. Simultaneous in-phase excitation of both sites (atoms) evidences superradiance, accelerating the radiation dynamics and significantly increasing the energy radiated into the lattice, reducing the energy remaining within the initially excited outlying sites. We also study the effect of applying an arbitrary phase difference to the optically excited sites, where we evidence subradiant dynamics for an out-of-phase input condition. In this case, the input excitation coincides with the profile of a bound state into the continuum~\cite{bic1,bic2,bic3,bic4,bic5,bic6}, and the energy remains almost perfectly trapped between the outlying sites. We effectively switch the dynamics into well-defined spatial states by varying the input amplitude or phase between the optical excitations. Our results draw inspiration from experimentally studying collective effects in quantum optics to use the phenomena for transport control and all-optical switching in photonic lattices.

\section{\label{theory}Theory and simulations}

\subsection{\label{model}Lattice model}

Our photonic kagome lattice under study consist of an array of single-mode optical waveguides which evanescently couple to their nearest-neighbors. The dynamic is well described by a Discrete Lineal Schr\"odinger Equation (DLSE)~\cite{rep1} that reads, in a general and compact form, as

\begin{equation}
-i\frac{\partial u_{\vec{n}}}{\partial z}=\sum_{\vec{m}} V_{\vec{n},\vec{m}}u_{\vec{m}}\ .\label{dls}
\end{equation}

\noindent Here, $u_{\vec{n}}$ describes the mode amplitude at the $\vec{n}$-site and $z$ is the propagation coordinate along the waveguides (which corresponds to time in quantum mechanics). $V_{\vec{n},\vec{m}}$ are the matrix coefficients defining the coupling interaction in between the nearest-neighbor sites $\vec{n}$ and $\vec{m}$, under the lattice geometry sketched in Fig.~\ref{fig1}(a), with horizontal $V_h$ and diagonal $V_d$ coupling constants. A kagome lattice~\cite{kag1,kag2} has three sites per unit cell, as shown by sites $A$, $B$ and $C$ in the same figure. The total power $P_{total}\equiv\sum_{\vec{n}} P_{\vec{n}}$ is a conserved quantity of model (\ref{dls}), with $P_{\vec{n}}\equiv|u_{\vec{n}}|^2$ the $\vec{n}$-th lattice site power.

We obtain the bands of the system by inserting into the model (\ref{dls}) a standard plane-wave (Bloch) ansatz of the form $u_{\vec{n}}(z)=\{A_0,B_0,C_0\} \exp(i\vec{k}_{\bot}\cdot\vec{r})\exp(ik_z z)$. Here, $A_0,B_0,C_0$ correspond to the amplitudes at the respective unit cell sites, $\vec{k}_{\bot}$ to the transversal wavevector, and $\vec{r}$ to a generic lattice position. $k_z$ corresponds to the longitudinal propagation constant or spatial frequency along the propagation direction $z$ (in a solid-state context, $k_z$ corresponds to the energy~\cite{rep1}). The linear spectrum is composed of three bands, as shown in Fig.~\ref{fig1}(b). Two of them are dispersive and connected by Dirac cones at the vertices of a hexagonal Brillouin zone. Dispersive bands are composed of extended propagating modes responsible for the transport on a given lattice system~\cite{rep1,kittel}. In our case, the third (lower) band at the bottom is quasi flat~\cite{repFB}. This third band becomes perfectly flat (i.e., $k_z=$ constant) only if all coupling constants are equal on a kagome geometry~\cite{kag1,kag2}; i.e., for a completely isotropic lattice. However, when the band is nearly (quasi) flat, their modes are very slow (massive) and do not contribute efficiently to energy transport.

\subsection{\label{sec:level1}Analogy to radiation}

A single outlying lattice site, which is evanescently coupled to a lattice array, can be considered as a quantum emitter coupled to a quasi-continuum structured reservoir \cite{Lambropoulos2000}. As long as the quantum system remains in the single-excitation regime, the same equations of motion describe the evolution of both systems. When the site/atom is initially excited, its excitation will decay into the array/reservoir in a process resembling radiation. The radiation behavior depends on the ratio in between the coupling $g$ of the single site to the array and the coupling $V$ between sites within the array. In the limit of weak coupling, $g/V\ll1$, the excited sites decay exponentially. In the strong coupling regime, $g/V\gg1$, the excitation is localized and oscillates between the outlying site and the nearest sites in the array (with a dimer-like effective dynamics). The behavior in the intermediate regime, $g/V\sim1$, is more complicated and depends strongly on the structure of the reservoir. Generally, the radiation begins as an exponential decay until reaching an approximately constant value that decays polynomially slowly at longer times\cite{Lambropoulos2000}. This general behavior even holds for atoms radiating into free space \cite{nonexpodecay}, where a pure exponential decay gives a good approximation.

In the weak coupling approximation, the mostly exponential decay depends on the coupling to the array $g$ and a finite density of states (DOS) \cite{Gonzalez-Tudela2017}. Fermi's golden rule tells us that the exponential decay rate is $\gamma=2\pi g^2 \rho (\Delta)$, where $\rho(\Delta)$ is the DOS at a frequency $\Delta$.  In the case of a waveguide coupled to a kagome lattice, the non-zero DOS at zero energy guarantees the excitation transport through the array. For two outlying sites radiating into the lattice array, their relative amplitudes and phases can lead to destructive or constructive interference. The case of constructive (destructive) interference enhances (suppresses) the radiation into the quasi-continuum, in analogy to the collective effects of superradiance (subradiance). A decay rate $\gamma$ could be collectively enhanced (suppressed) to reach a decay rate $\gamma_{\rm{tot}}=2 \gamma$ ($\gamma_{\rm{tot}}=0$). Collective effects of radiation into 2D structured reservoirs have been theoretically studied \cite{Gonzalez-Tudela2017,Gonzalez-Tudela2017b,Yu2019}, but to our knowledge, they lack experimental implementations.


\subsection{\label{sec:level2}Dynamical analysis}

We numerically integrate the model (\ref{dls}) to study the radiation phenomena in the waveguide array, establishing an analogy where a single outlying site acts as an atom and the lattice acts as a continuum reservoir~\cite{longhi20}. Exciting the system on a single outlying site allows studying standard radiation processes, while exciting two sites simulates collective behaviors. Fig.~\ref{fig1}(a) shows the $A$ and the $C$ outlying sites acting as radiating atoms, as emphasized by a green ellipse. Both sites connect to the rest of the lattice through a $B$ site. In this scheme, we can use the analogy of atoms radiating into a 2D kagome lattice and study its dependence under different input conditions. To gain insight into the dynamical properties, as well as to approach to the experimental regime, we numerically study the isotropic ($V_d=V_h$) and weak anisotropic ($V_d/V_h=1.2$) cases. We characterize the dynamics by computing the remaining power at the isolated $A$ and $C$ atomic sites ($P_{\rm{atoms}}$), both located at the right-upper lattice corner, and dividing it by the total power in system ($P_{\rm{total}}$), including these atoms. We define $P_{\rm{atoms}}/P_{\rm{total}}$ in analogy to the atomic excitation probability to quantify the radiation process and the dynamics of the system. 

\begin{figure}[t!]
\centering
\includegraphics[width=1\linewidth]{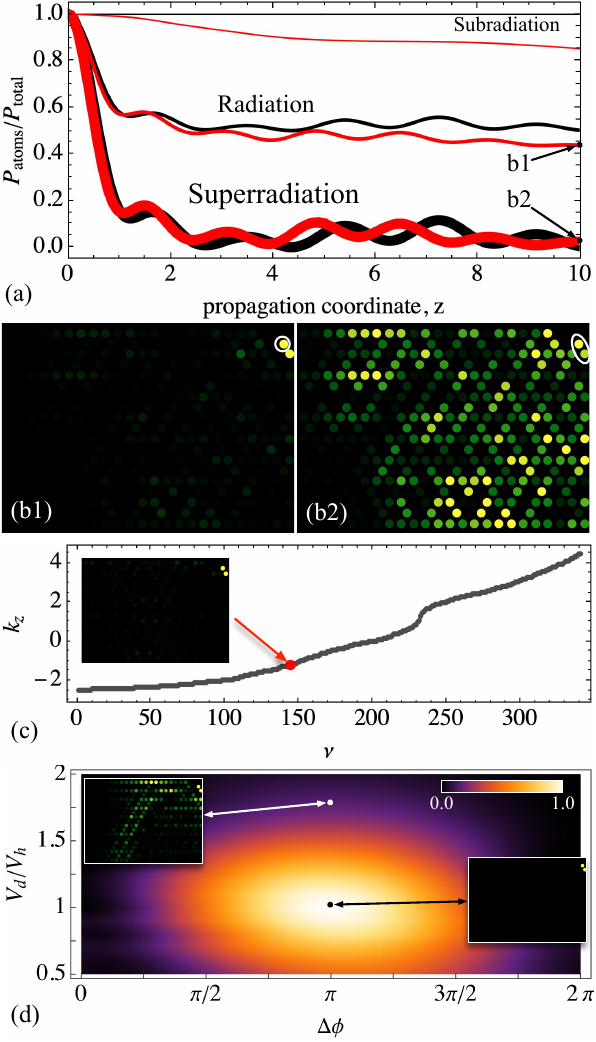}
\caption{(a) $P_{\rm{atoms}}/P_{\rm{total}}$ versus propagation coordinate $z$ for isotropic (black) and anisotropic (red) kagome lattices. The normal lines show the dynamics after optically exciting a single atomic site. In contrast, thicker and thiner lines show the dynamics of two atomic sites optically excited in and out-of-phase, respectively. (b1) and (b2) Output intensity profiles at $z=10$ cm for single-site and in-phase double-site excitations as indicated in (a). (c) Kagome spectrum for the finite lattice having 343 sites and $V_d/V_h=1.2$. The inset shows the respective edge state in the continuum. (d) $P_{\rm{atoms}}/P_{\rm{total}}$ versus $\Delta\phi$ and $V_d/V_h$, for the excitation of two atoms after a propagation of $z=L=10$ cm. Insets show the indicated cases.}
\label{fig2}
\end{figure}

Figure~\ref{fig2}(a) presents a compilation of our numerical results for isotropic (black) and anisotropic (red) lattices. We first excite a single waveguide and study the power evolution at the atomic site. We observe (normal lines) a similar behavior for both lattice cases, with approximately one-half of the energy being radiated to the lattice and the other half oscillating at the region of the atomic sites. As all the coupling constants are of the same order in our lattice, we assume this observation corresponds to an intermediate radiation regime~\cite{Crespi2019}, where the energy is shared between the two atoms and the lattice. Fig.~\ref{fig2}(b1) shows the output profile after a propagation length $L=10$ cm for the anisotropic case. We observe that both atoms are strongly excited, with an almost equal intensity. These two atoms create a dimer-like system, generating oscillation between them, while simultaneously the light is been radiated efficiently to the rest of the lattice. However, this is not so evident in Fig.~\ref{fig2}(b1) due to the large intensity differences in between the atomic and lattice sites (the energy is homogeneously distributed into the lattice, with a low-intensity density per site of $\sim0.001$ compared to $\sim0.25$ contained at each atom).

A collective superradiant effect occurs when the two atoms are simultaneously excited in phase. Thicker lines in Fig.~\ref{fig2}(a) show similar dynamics for both lattices, where we observe a quite notorious enhanced radiative dynamics. We observe a faster energy transport into the lattice, where for $z\approx0.5$ cm around $50\%$ of the energy has been already disseminated (for a single atomic site excitation, this occurs at $z\approx2$ cm). However, even more important, almost all the energy has been disseminated to the lattice for $z\approx2.5$ cm. This figure shows a noticeable and robust difference between the regimes of radiation and superradiation for this 2D kagome lattice. Fig.~\ref{fig2}(b2) shows the output profile for this case, at the propagation length of $z=L$, where we observe a strong contrast with the single atomic site excitation shown in Fig.~\ref{fig2}(b1). This numerical observation clearly shows that the chosen kagome configuration constitutes an excellent scenario for radiative-like studies. 

Now, we study the effect of considering a simultaneous excitation of both atoms but having a nontrivial input phase structure. This idea comes from a recent work where authors use a Lieb ribbon photonic lattice~\cite{filtercontrol} to study the excitation of $0$- and $\pi$-phase qubits. Taking advantage of the FB properties of a Lieb geometry, those authors could cancel the transport to the lattice for an out-of-phase excitation. On the other hand, the energy radiates through the system for an in-phase condition. In our case, the lattice anisotropy demands us to use a balanced amplitude condition to fully cancel the transport through the lattice while exciting both atoms in an out-of-phase configuration~\cite{sawtooth}. Suppose we define the amplitude of the isolated atoms as $a$ and $c$, top and right, respectively. We should satisfy the condition $V_d a+V_h c=0$ to achieve the required destructive interference at the connector site $B$. In this case, the transport through the lattice would be minimal, with most of the energy remaining localized at both atomic sites with $P_{atoms}/P_{total}\approx1$. Thiner line plots in Fig.~\ref{fig2}(a) show this regime for both lattice cases. We observe how the energy remains trapped only at the atomic sites for a perfectly isotropic lattice, while for a weakly anisotropic configuration, the energy slowly leaks into the lattice. The out-of-phase excitation relates to a compact stationary state, which may correspond to a bound edge state in the continuum~\cite{bic1,bic2,bic3,bic4,bic5,bic6}. This state has two sites different from zero, only for the isotropic case $V_d=V_h$. However, for anisotropic lattices, the energy slowly radiates into the bulk, but it appears as an effective localized state for short propagation distances. 

Figure~\ref{fig2}(c) shows the eigenvalue spectra for the finite lattice structure under study [see Fig.~\ref{fig3}(b)] considering $V_d/V_h=1.2$. We observe that there is a state (red dot) inside the second band, at a frequency $k_z\approx -1.2$, which is highly trapped at the atomic sites region, as the intensity profile shows at the inset. In fact, this is the highest localized state for this lattice geometry, with a participation number~\cite{kag1,kag2} of $5.3$ for $V_d/V_h=1.2$. The intensity ratio between both atomic sites is $1.23$, and the rest of the lattice amplitude is minimal but not zero. In fact, by using an out-of-phase input condition, we numerically find that $\sim96\%$ of the energy is trapped at the atomic sites after propagation distance of $z=10$ cm. On the other hand, for an in-phase atoms excitation, this value strongly decreases to $\sim2\%$. 

To characterize this better, we run several simulations by varying the lattice anisotropy $V_d/V_h$ and the input phase $\Delta\phi$ in between two equal amplitude atomic sites. After running each simulation, up to a propagation distance of $z=10$ cm, we compute the energy remaining at atomic sites ($P_{\rm{atoms}}/P_{\rm{total}}$) and show our compiled results in Fig.~\ref{fig2}(d). There, we observe an evident optical switch effect, which could be fully controllable by external optical means~\cite{filtercontrol}. We notice that, for a perfect isotropic regime ($V_d=V_h$) and an out-of-phase ($\Delta\phi=\pi$) input condition, the energy remains perfectly trapped at the atomic sites, clearly shown at the right-panel inset. Around this parameter region, $P_{atoms}/P_{total}\approx1$ due to the effective excitation of a bound edge state in the continuum~\cite{bic1,bic2,bic3,bic4,bic5,bic6}, which has much larger amplitudes at the $A$ and $C$ atomic sites. Therefore, this input condition effectively excites a localized state at the atomic sites, which naturally does not radiate, or it only does so weakly. This regime is a perfect analogy to the subradiant regime in quantum optics~\cite{oriol22}. On the other hand, for an in-phase input excitation ($\Delta\phi\approx0,2\pi$), the energy is fully superradiated into the lattice, independently of the lattice anisotropy, as Figs.~\ref{fig2}(a) and (b2) show. We also notice [see the left-panel inset in Fig.~\ref{fig2}(d)] that for an out-of-phase input condition, on a highly anisotropic lattice, the energy is also well radiated into the bulk. This effect is due to the absence of compact edge states to excite at the atomic sites, with only propagating modes available in the lattice after excitation. A larger anisotropy effectively decouples the atomic site $C$ from the connector site $B$, and no localized edge state is longer possible at the atomic sites region.

\section{Experimental excitation}

\begin{figure}[t!]
\centering
\includegraphics[width=1\linewidth]{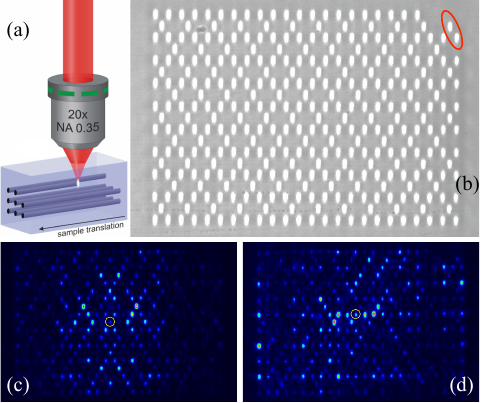}
\caption{(a) fs writing technique. (b) A fs written kagome photonic lattice, including the effective site atoms emphasized by a red ellipse. (c) and (d) Ouput intensity experimental images after a $C$ and $B$ bulk site excitation, respectively.}
\label{fig3}
\end{figure}


The kagome lattice under study was fabricated using the direct femtosecond laser-writing technique~\cite{Alex1} [see the sketch in Fig.~\ref{fig3}(a)] on a $L=10$ cm-long fused silica glass wafer. Fig.~\ref{fig3}(b) shows a microscope image of our fabricated kagome lattice with $343$ single-mode waveguides (at $633$ nm), having a lattice -- center to center -- spacing of $20\ \mu$m; i.e., a geometrically isotropic configuration. However, the waveguide ellipticity becomes quite evident after white light illumination, with an effective profile of $\sim4\times 12\ \mu$m~\cite{Alex1}. This ellipticity affects the propagation dynamics on this lattice due to the different evanescent coupling among different waveguides, depending on the waveguide orientation. Specifically, in this case, the horizontal coupling constant $V_h$ becomes smaller than the diagonal one ($V_d$) at an equal geometrical distance. This asymmetry implies that our perfectly symmetric lattice configuration becomes effectively anisotropic in terms of dynamical properties.

First of all, we study this lattice experimentally using a standard characterization setup, which consists on focusing and linearly polarizing a HeNe laser beam to excite individual bulk waveguides. Figs.~\ref{fig3}(c) and (d) show discrete diffraction patterns at the output facet for $C$ and $B$ bulk excitations, respectively. Both cases show excellent transport properties with the light fully exploring the lattice. The $C$-site excitation shows a more vertically oriented pattern due to the first hopping with the up and down $A$ and $B$ sites. On the other hand, a $B$-site excitation shows a more horizontal distribution of the energy through the lattice, with some weak localization tendency in the surroundings of the input excitation. This could be due to a better excitation of the quasi-flat band formed by slow propagating modes. Nevertheless, in this case, the light explores quite well the lattice as it can be noticed by observing some localized patterns at the lattice surface [see Fig.~\ref{fig3}(d)]. 

\begin{figure}[t!]
\centering
\includegraphics[width=1\linewidth]{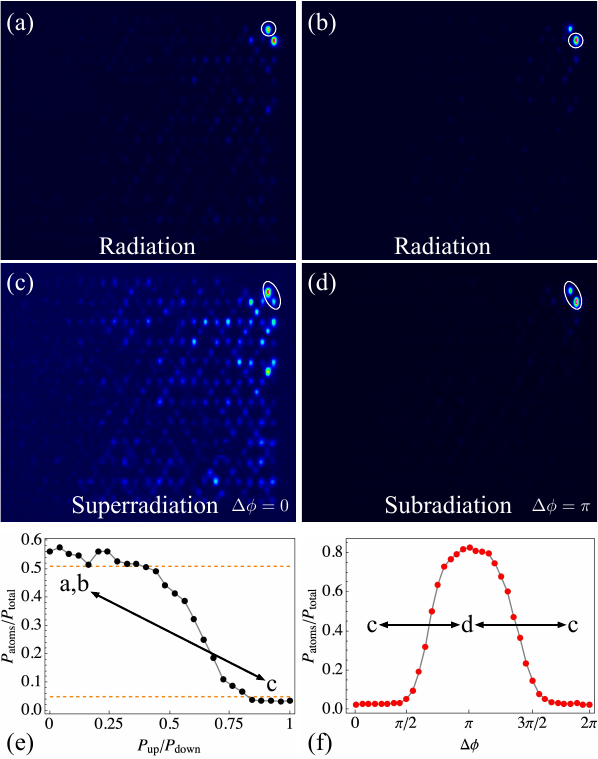}
\caption{(a) and (b) Output intensity profiles for up ($C$) and down ($A$) atomic site excitations, respectively. (c) and (d) Output intensity profiles for a simultaneous $A$ and $C$ in-phase and out-of-phase atoms excitation, respectively. Yellow circles show the corresponding input positions. (e) and (f) $P_{atoms}/P_{total}$ vs $P_{up}/P_{down}$ and $\Delta\phi$, respectively. Letters indicate the relation with panels (a), (b), (c) and (d). The experimental data was measured for $640$ nm on a $L=10$ cm-long kagome photonic lattice.}
\label{fig4}
\end{figure}

Now, we implement an image setup based on a sequence of two spatial light modulators (SLMs)~\cite{ABPRL}. In the first stage, we use a transmission SLM to modulate the amplitude of a $640$ nm wide laser beam and generate one or two light disks to excite one or two atoms, respectively. In the second stage, we use a reflective SLM to add a phase pattern to the generated amplitude modulated profile. In this way, we can simultaneously excite one or more waveguides with a well-defined amplitude and phase structure. We first excite every atom independently and observe the differences in the fabricated kagome lattice. Figures~\ref{fig4}(a) and (b) show the excitation of the upper $C$ and bottom $A$ isolated atomic sites. The experiments show that the upper atomic site excitation radiates energy through the lattice more efficiently than the bottom atomic site. Nevertheless, both cases show a slow radiation process with an amount of radiated energy around $50\%$, as expected from the numerical simulations shown in Fig.~\ref{fig2}(a). [As the experimental figures are normalized to the maximum intensity, the lattice background looks very weak, similar to the simulation shown previously in Fig.~\ref{fig2}]

Figure~\ref{fig4}(c) shows the collective effect of superradiance when both atoms are excited in phase ($\Delta\phi=0$), with both constructively radiating to the lattice bulk. We observe a well-disseminated output pattern, with the light exploring the lattice freely and with less than $5\%$ of the total power remaining at the atomic sites. Although the intensity looks higher at those sites, the additive contribution of the lattice sites is indeed much higher. The contrast in between independent atomic radiation and superradiation phenomena on our kagome structure is quite evident by a simple eye inspection of these experimental images.

On the other hand, by adding a $\pi$ phase difference between both excited atoms, we induce destructive interference dynamics at the connector $B$ site. This interference produces that the energy radiated to the lattice, at the experimental dynamical scale of $L=10$ cm, is around $15 \%$. Therefore, this input condition excites an almost perfect compact localized edge state, which remains trapped at the excitation region with a slow leaking into the lattice. This result is in very good agreement with the numerical results presented in Fig.~\ref{fig2}.
 
Now, we run a more intensive set of experiments, taking advantage of the possibilities of our image setup configuration. Specifically, we first set the excitation phase difference between both atomic sites as zero and we only vary the amplitude at the upper ($C$) atomic site, while keeping constant the amplitude at the bottom ($A$) one. This way, we can experimentally study the dynamic transition between pure radiative and superradiant processes. We show our collected results in Fig.~\ref{fig4}(e), where we observe a well-defined transition between these two regimes, with the letters indicating the panels at the same figure. These two clear regimes, with two well-defined plateaus, can be used as an optical switch. By controlling the radiance and superradiance properties on our kagome lattice, we can transit from a weakly radiated pattern into a strongly radiated one and decide, in a very controllable way, the radiation state we need; i.e., a photonic amplitude valve/switch. 

Finally, using the same image setup, we implement an experiment where we excite both atoms simultaneously with the same amplitude, but now by applying a controlled phase difference $\Delta \phi$ between the two atoms. Fig.~\ref{fig4}(f) shows our compiled results where we observe an almost perfect phase-controlled all-optical switch. There are well-defined states with the energy transiting from a superradiative pattern (with almost no energy at the atomic sites) at $\Delta \phi=0,2\pi$ into a subradiative one at $\Delta \phi=\pi$ (with most of the energy remaining trapped at atomic sites). In this case, we can select two very different dynamical states with high experimental precision by just controlling the phase difference between the atoms~\cite{filtercontrol}. Both experiments show a clear opportunity to use the radiative processes of a given lattice structure to externally control different output spatial patterns on demand and to use them as, for example, state logical bits to transmit optical information. 

\section{Conclusions}

In this work, we use a photonic kagome lattice to numerically simulate and experimentally demonstrate collective radiative phenomena in structured two-dimensional systems, presenting a precise all-optical control over these processes. The experiments demonstrate the transition between radiative, superradiative, and subradiative processes, showcasing the potential for optical switching and transport control in lattice arrays. An in-phase excitation of two outlying sites/atoms yields superradiance through a kagome lattice array, which accelerates radiation dynamics and significantly enhances the energy radiated to the lattice. In contrast, an out-of-phase excitation leads to subradiant dynamics, wherein energy remains highly confined between the excited atomic sites.

The study advances our knowledge of simulating quantum optical phenomena within photonic lattices and highlights the practical utility of these effects. These findings lay the foundation for future exploration of simulating quantum optical effects in two-dimensional structured reservoirs, setting the stage for harnessing these phenomena in photonic systems. The research contributes to the burgeoning field of quantum optics and photonic lattices, where manipulating light and its quantum properties could impact various technologies and applications.

\begin{acknowledgments}
This work was supported by FONDECYT grants 1231313 and 11200192, CONICYT-PAI grant 77190033. P.S. is a CIFAR Azrieli Global Scholar in the Quantum Information Science Program. A.S. acknowledges funding from the Deutsche Forschungsgemeinschaft (grants SZ 276/9-2, SZ 276/19-1, SZ 276/20-1, SZ 276/21-1, SZ 276/27-1, GRK 2676/1-2023 ‘Imaging of Quantum Systems’, project no. 437567992, and SFB 1477 ‘Light–Matter Interactions at Interfaces’, project no. 441234705).
\end{acknowledgments}

\section*{Author declarations}
\subsection*{Conflict of Interest}
The authors have no conflicts to disclose.
\subsection*{Author Contributions}
\textbf{Ignacio Salinas: Investigation, Formal Analysis}. \textbf{Javier Cubillos: Data curation, Formal Analysis, Investigation}. \textbf{Alexander Szameit: Investigation, Funding acquisition}. \textbf{Pablo Solano: Formal Analysis, Funding acquisition, Writing}. \textbf{Rodrigo A. Vicencio: Formal Analysis, Funding acquisition, Investigation, Methodology, Resources, Supervision, Visualization, Writing}.

\section*{Data Availability Statement}

The data that support the findings of this study are available from the corresponding author upon reasonable request.

\nocite{*}
\bibliography{aipsamp}

\end{document}